\documentclass[12pt]{article}

\usepackage{epsfig}

\begin{document}

\title{New Universality Class in three dimensions:\\
                the Antiferromagnetic $RP^2$ model}

\author{H.~G.~Ballesteros, L.~A.~Fern\'andez, \\
        V.~Mart\'{\i}n-Mayor and A.~Mu\~noz Sudupe\\
\it Departamento de F\'{\i}sica Te\'orica I, Facultad de CC. F\'{\i}sicas,\\
\it Universidad Complutense de Madrid, 28040 Madrid, Spain.\\
{\small e-mail: {\tt hector,laf,victor,sudupe@lattice.fis.ucm.es}}}

\date{November 2, 1995}

\maketitle

\begin{abstract}
We present the results of a Monte Carlo simulation of the $RP^2$ model
in three dimensions with negative coupling.  We observe a second order
phase transition between the disordered phase and an
antiferromagnetic, unfrustrated, ordered one.  We measure, with a
Finite Size Scaling analysis, the thermal exponent, obtaining
$\nu=0.784(8)$. We have found two magnetic-type relevant operators
whose related $\eta$ exponents are $0.038(2)$ and $1.338(8)$
respectively.
\end{abstract}

\newpage

The theory of Critical Phenomena offers a common framework to study
problems in Condensed Matter Physics (CMP) and in High Energy Physics
(HEP). In both areas, the concepts of Spontaneous Symmetry Breaking
(SSB) and of Universality allow to relate problems in principle very
different.

The usual Heisenberg model, associated with the standard ferromagnetic
Non Linear $\sigma$ Model (NL$\sigma$M), has a SSB pattern of type
$SO(3)/SO(2)$.  With the introduction of nontrivial AntiFerromagnetic
(AF) interactions the SSB pattern normally changes completely and,
usually, frustration is generated.  In particular, a SSB pattern shared
by several AF models is $SO(3)\times SO(2)/SO(2)$.  For instance, some
frustrated quantum AF Heisenberg models~\cite{DOMBRE}, or the
Helimagnets and Canted spin systems~\cite{HELICAL,AZARIA} are examples
of this behavior.  Frustrated quantum spin models are specially
interesting because of their possible relation with High Temperature
Superconductivity~\cite{HTSC}.

As a general consequence of the Weinberg Theorem the low energy
physics of a system is completely determined by its SSB pattern, the
effective Lagrangian for the system being the corresponding
NL$\sigma$M.  In this framework a study has been carried out for
\hbox{$SO(3)\times SO(2)/SO(2)$} in perturbation theory~\cite{AZARIA},
where the main conclusion reached is that the only possible
nontrivial critical point in three dimensions is that of $O(4)$.  In
spite of that, we have found in a nonperturbative lattice formulation
of the same model, a critical point with exponents clearly different
from those of $O(4)$.

{}From the HEP point of view, it is of great interest to understand
whether AF interactions can generate new Universality Classes.  One
could even hope that nontrivial antiferromagnetism would be the
ingredient needed in order to nonperturbatively formulate interacting
theories in four dimensions.

In a previous work~\cite{O3A} we found that, on a three dimensional AF
$O(3)$ model, the only new phase transitions generated were first
order.  We will consider in this letter the $RP^2\equiv S^2/Z_2$ (real
projective space) spin model  in three dimensions. We
place the spins on a cubic lattice with a nearest neighbors
interaction:
\begin{equation}
S= \beta \sum_{<i j>} ({\mbox{\boldmath $v$}}_i\cdot \mbox{\boldmath
$v$}_j)^2 ,
\label{ACCION}
\end{equation}
where $\{\mbox{\boldmath $v$}_i\}$ are normalized real three-components
vectors.  The local $Z_2$ symmetry $\mbox{\boldmath $v$}_i\to
-\mbox{\boldmath $v$}_i$ is preserved even after the SSB (Elitzur's
theorem), and so, the sense of a spin is irrelevant, it is only its
direction that matters.  It is not hard to see~\cite{RP2LARGO}, that
in the AF case (\ref{ACCION}) is a lattice discretization of the
action $\int {\rm tr}[P{(R^{-1}\partial_{\mu}R)}^{2}]$, where $R\in
SO(3)$, and $P$ is the diagonal matrix $\{g,g,-g\}$, with $g$ being
the coupling.  This is just a particular case of the NL$\sigma$M
considered in reference~\cite{AZARIA}.

For $\beta$ positive this model presents a weak first order phase
transition which has been used to describe liquid
crystals~\cite{LIQUIDCRYS}. The ordered phase corresponds to states
where all spins are aligned.

For $\beta$ negative there is also an AF ordered phase with a more
complex structure~\cite{SHROCK}. There is a second order phase
transition between the disordered phase and an ordered AF one.  Let us
call a site even (odd) when the sum of its coordinates $x+y+z$ is even
(odd). A state where, for instance, all spins on even sites are
aligned in a given direction, and those on odd sites lie randomly in
the orthogonal plane, has zero energy.  So, at $T=0$ the ground state
is highly degenerate with a global $O(2)$ symmetry.  However, when
fluctuations are taken into account, it can be shown in the low
temperature limit~\cite{RP2LARGO},
that both sublattices are aligned in mutually orthogonal directions,
as a consequence of Villain's {\it order from disorder}
mechanism~\cite{VILLAIN}.
We remark that the remaining $O(2)$ symmetry is broken.
We will show, with Monte Carlo simulations,
that the breaking also holds in the critical region.

In order to discuss the observables measured, let us construct the
(traceless) tensorial field $\bf T$ with components
\begin{equation}
{\rm T}_i^{\alpha\beta}= v_i^\alpha v_i^\beta-
        \frac{1}{3}\delta^{\alpha\beta} ,
\label{TENSORFIELD}
\end{equation}
and its Fourier Transform $\widehat{\bf T}$ in a $L\times L \times L$
lattice with periodic boundary conditions.

The intensive staggered (non staggered) magnetization can be defined
in terms of the tensorial field as the sum of the spins on even sites
minus (plus) those on odd sites, or equivalently
\begin{equation}
{\bf M}_s=\frac{1}{V}\widehat{\bf T}_{(\pi,\pi,\pi)} \quad,\quad ({\bf
M}=\frac{1}{V}\widehat{\bf T}_{(0,0,0)}),
\label{MAGS}
\end{equation}
where $V$ is the lattice volume. We have observed a phase transition
at $\beta\sim -2.41$ for which ${\bf M}_s$ is an order parameter (zero
value in the disordered phase and a clear nonzero value in the
$L\to\infty$ limit in the ordered one). The magnetization ${\bf M}$ is
also an order parameter.  As these operators correspond to different
irreducible representations of the translations group, we will study
the scaling properties of each observable independently.

To measure in a Monte Carlo simulation on a finite lattice we have
constructed scalars under the $O(3)$ group. For the magnetization and
the susceptibility we compute respectively
\begin{equation}
M=\left\langle \sqrt{{\rm tr} {\bf M}^2}\right\rangle\quad,\quad
\chi=V \left\langle {\rm tr}{\bf M}^2\right\rangle ,
\end{equation}
and analogously for the staggered observables.

We have also measured the {\em second momentum} correlation length,
which is expected to have the same scaling behavior at the critical
point as the exponential (physical) one, but it is much easier to
measure~\cite{CORREL}
\begin{equation}
\xi=\left(\frac{\chi/F-1}{4\sin^2(\pi/L)}\right)^{1/2},
\end{equation}
where $F$ is the mean value of the trace of $\widehat{\bf T}$ squared
at minimal momentum ($2 \pi/L$ in direction $x$, $y$ or $z$).  To
define $\xi_s$ we use $\chi_s$ and compute $F_s$ from $\widehat{\bf
T}$ at momentum $(2 \pi/L+\pi,\pi,\pi)$ and permutations.

The action (\ref{ACCION}) is suitable for cluster update methods by
using the Wolff's embedding algorithm \cite{WOLFF}.  We have checked
the performance of both the Swendsen-Wang \cite{SWENDWANG} and the
Single Cluster methods~\cite{WOLFF}. Unfortunately, due to the AF
character of the interaction, the critical slowing down is not reduced
in any case, as there is always a large cluster that contains most of
the lattice sites.  In fact, we have measured a dynamic exponent
$z\approx 2$ for both methods.

We have also developed a Metropolis algorithm. Near the transition the
spin fluctuations are large, and a spin proposal uniformly distributed
over the sphere is accepted with nearly 30\% probability. So we have
used a 3 hits algorithm, reaching a mean 70\% acceptance.

Regarding the performance of the three methods mentioned for a given
lattice size, the differences are very small in terms of the CPU
time. We have selected the Metropolis method which is slightly faster.

\begin{table}[ht]
\begin{center}
\begin{tabular}{|r|c|c|c|c|}\hline
$L$ & MC sweeps($\times 10^6$)
& $\tau_{\chi_s}$ &$\tau_\chi$ &$\tau_E$\\\hline\hline
6       & \hphantom{0}6.71   &7.4   &5.8    &0.60            \\ \hline
8       &            17.07   &11.4  &7.4    &0.73            \\ \hline
12      & \hphantom{0}6.51   &24.8  &12.8   &1.02            \\ \hline
16      &            22.14   &44.1  &21.4   &1.30            \\ \hline
24      & \hphantom{0}8.77   &107.  &48.    &1.82            \\ \hline
32      &            10.13   &179.  &87.    &2.27            \\ \hline
48      & \hphantom{0}3.93   &430.  &205.   &3.10            \\ \hline
\end{tabular}
\caption{Number of Monte Carlo sweeps performed for different lattice
sizes. Measures have been taken every 10 sweeps. The integrated
autocorrelation times (in sweeps) for both magnetizations and for the
energy are also displayed. The statistical errors are below the 5\%
level. We have discarded in each case about $200\tau_{\chi_s}$
iterations for thermalization.}
\label{SIMU}
\end{center}
\end{table}

In table \ref{SIMU} we display the number of Monte Carlo sweeps
performed for the different lattice sizes as well as the integrated
autocorrelation times for the observables $\chi$, $\chi_s$ and the
energy. The total CPU time has been the equivalent of 12 months of DEC
Alpha AXP3000 distributed over several workstations.

Every 10 Monte Carlo sweeps we store individual measures of the energy
and of the Fourier transform of the tensorial magnetization at
suitable momentum values.  We have used the spectral density
method~\cite{FERRSWEN} to extrapolate in a neighborhood of the
critical point.  The data presented here correspond to simulations at
two $\beta$ values ($-2.41$ and $-2.4$).  We compute the quantities
referred above as well as their $\beta$-derivatives through the
connected correlations with the energy.

We have firstly analyzed several quantities that present a peak near
the transition point. As we have found that the specific heat does not
diverge, we have to limit ourselves to study quantities related with
magnetization operators.  The advantage of measuring a peak height is
that its position also defines an apparent critical point allowing for
a very simple and accurate measure.  Unfortunately, quantities like
the $\beta$-derivatives of the magnetizations or the connected
susceptibilities \hbox{($\chi_{\rm con}\equiv\chi-VM^2$)} present
their peaks far away from the critical point suffering from large
corrections to scaling.  For example, $\chi_{\rm con}$ in the
$L=16$ lattice peaks at $\beta=-2.29$ where the correlation length
$\xi$ is one half of its value at the critical point.  This is due to
the weakness of the tensorial ordering: the staggered magnetization,
for instance, does not reach one half of its maximum until
$\beta<-3.5$.

Another possibility is to obtain the infinite volume critical point by
other means and, then, to measure the different quantities at this
point. By studying the matching of the Binder parameter for the
staggered magnetization $V_{M_s}=1- \langle({\rm tr}{\bf
M}_s^2)^2\rangle/(3\langle {\rm tr}{\bf M}_s^2\rangle^2)$, as well as
that of $\xi_s(L,\beta)/L$ and the corresponding non staggered
quantities, we conclude that
\begin{equation}
\beta_c \in [-2.415,-2,405]\ .
\label{BETACRI}
\end{equation}
To improve the above determination it is necessary a careful
consideration of the corrections to scaling. This subject will be
discussed in a forthcoming paper~\cite{RP2LARGO}.

\begin{figure}[t]
\begin{center}
\epsfig{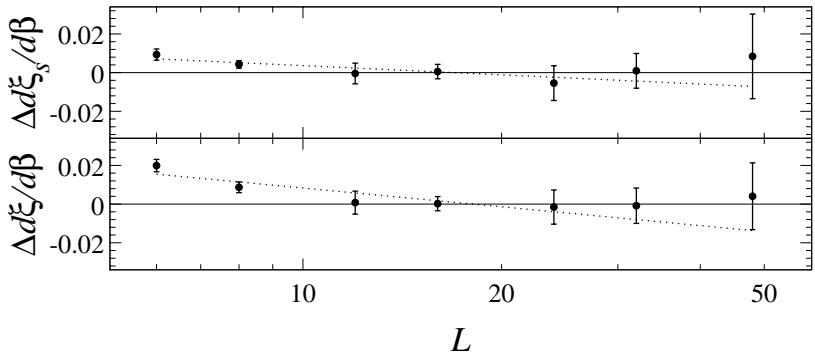}
\caption{Deviations from a power law fit of the $\beta$-derivative of
the staggered (upper side) and non staggered (lower side) correlation
lengths at the critical point, using data from lattice sizes with
$L\ge 16$.  The dotted lines correspond to a fit using all data
sizes.}
\label{LOGLOG}
\end{center}
\end{figure}

In the case of quantities that change rapidly at the critical
point as the magnetizations do, the errors in the determination of the
critical point affect very much the results, and this method is not
accurate. Nevertheless, we have found important quantities, like
the $\beta$-derivatives of the correlation lengths, which are very
stable. Both $d\xi_s/d\beta$ and $d \xi/d\beta$ should scale as
$L^{1+1/\nu}$. Fitting the data from all lattice sizes we obtain an
acceptable fit: $\nu=0.793(2)$ with $\chi^2/{\rm dof}=2.0/5$ and
$\nu=0.787(2)$ with $\chi^2/{\rm dof}=5.5/5$; however, if we discard
the $L=6$ data the fitted parameters change significantly and change
again after discarding the $L=8$ ones. The fits for $L\ge 12$ and
$L\ge 16$ agree within errors.  We, thus, choose the $L\ge 12$ data
for computing $\nu$ but take the statistical error from the fit with
$L\ge 16$ (see figure \ref{LOGLOG}):
\begin{equation}
\begin{array}{lll}
\xi_s&:  &\quad \nu=0.788(7)\\
\xi  &:  &\quad \nu=0.779(6)\ .
\label{FITS}
\end{array}
\end{equation}

To estimate the errors, not considered in (\ref{FITS}), associated
with the uncertainty in the determination of the critical point we
repeat the fits with $\beta$ at the limits of the interval
(\ref{BETACRI}). We observe that $\nu$ changes \hbox{by an amount of a
$1\%$}.

To avoid the problems reported above, we have also used a method
directly based on the Finite Size Scaling ansatz, that allows to write
the mean value of any operator $O$ as
\begin{equation}
\langle O(L,\beta) \rangle=L^x f_O(\xi(L,\beta)/L)+\ldots\ ,
\label{FSS}
\end{equation}
where $\xi(L,\beta)$ is the correlation length measured at coupling
$\beta$ in a size $L$ lattice, $f_O$ is a smooth operator-dependent
function and $x$ depends also on $O$. The dots stand for corrections
to scaling.

\begin{table}[t]
\begin{center}
\begin{tabular}{|r|l|l|l|l|l|l|}\hline
    & \multicolumn{2}{|c|}{$\nu$}
    & \multicolumn{2}{c|}{$\eta_s$}
    & \multicolumn{2}{c|}{$\eta$}      \\
\cline{2-3} \cline{4-5} \cline{6-7}
$L$ & \multicolumn{1}{|c|}{$d\xi_s/d\beta$}
    & \multicolumn{1}{c|}{$d\xi/d\beta$}
    & \multicolumn{1}{c|}{$\chi_s$}
    & \multicolumn{1}{c|}{$M_s$}
    & \multicolumn{1}{c|}{$\chi$}
    & \multicolumn{1}{c|}{$M$}\\\hline\hline
6  & 0.786(6) & 0.790(6)  & 0.0431(10)
        & 0.0474(9)  & 1.442(2) & 1.447(2) \\\hline
8  & 0.785(4) & 0.781(4)  & 0.0375(7)
        & 0.0409(8)  & 1.413(2) & 1.416(2) \\\hline
12 & 0.789(8) & 0.782(9)  & 0.0357(17)
        & 0.0382(18) & 1.391(3) & 1.393(3) \\\hline
16 & 0.787(9) & 0.781(8)  & 0.0371(19)
        & 0.0390(19) & 1.379(4) & 1.381(4) \\\hline
24 & 0.77(2)  & 0.77(2)   & 0.038(5)
        & 0.038(5)   & 1.362(8) & 1.365(9) \\\hline
\end{tabular}
\caption{Critical exponents obtained from a Finite Size Scaling
analysis using data from lattices of sizes $L$ and $2L$. In the second
row we show the operator used for each column.}
\label{EXPONENTES}
\end{center}
\end{table}

\begin{figure}[t]
\begin{center}
\epsfig{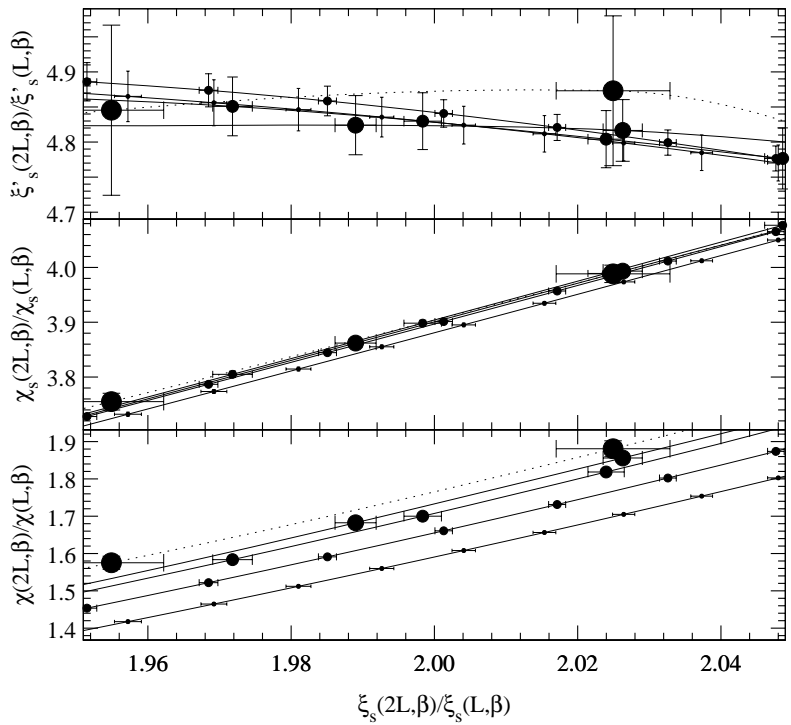}
\caption{Quotients of several observables as a function of the
quotient of the staggered correlation lengths. The sizes of the
symbols are a growing function of the lattice size. $\xi_s'$ is the
$\beta$-derivative of the staggered correlation length.}
\label{COCIENTES}
\end{center}
\end{figure}

Measuring $\langle O\rangle$ at the same coupling in lattices $2L$ and
$L$ and using (\ref{FSS}) we can write for their quotient
\begin{equation}
Q_O=2^x \frac{f_O\left(\frac{\xi(2L,\beta)}{2L}\right)}
       {f_O\left(\frac{\xi(L,\beta)}{L}\right)}+\ldots\ .
\end{equation}
Considering the dependence of $Q_O$ on
$\rho\equiv\xi(2L,\beta)/\xi(L,\beta)$, we just have to measure at the
point where $\rho=2$ to obtain $Q_O=2^x$ up to corrections to
scaling. $x$ can be written in terms of the critical exponents:
$x=\gamma/\nu$ for $\chi$, $x=-\beta/\nu$ for $M$, etc.. To compute
$\nu$ we can use $O=d\xi/d\beta$ for which $x=1+1/\nu$.  As the
quantity $\rho$ is an observable and not an external parameter, like
$\beta$, this procedure does not require a previous determination of
$\beta_c$.  Another advantage is that the result depends only on
measures in just two lattices what allows a better error estimation.
In columns 2 and 3 of table \ref{EXPONENTES} we report the results for
the thermal exponent $\nu$ obtained from data in lattices $L$ (first
column) and $2L$, using as operators the $\beta$-derivatives of the
correlation lengths $\xi_s$ and $\xi$ respectively. Even in the
smaller lattices we do not observe any corrections to scaling.  The
data from columns 2 and 3 are very correlated statistically and so, by
taking the mean value, the errors are only slightly reduced.  We
select as our best estimation the mean of the results for the lattices
16-32:
\begin{equation}
\nu=0.784(8).
\label{NU}
\end{equation}
To be compared with the values $\nu(O(3))=0.704(6)$~\cite{O3} and
$\nu(O(4))=0.748(9)$~\cite{O4}.

In the case of the magnetic exponents, due to the large slope of $Q$
as a function of $\rho$ (see figure \ref{COCIENTES}), a correct
determination of the errors requires to take into account the
statistical correlations of the whole data. We obtain for $\gamma_s$
and $\beta_s$, in the most favorable cases, errors as small as a
0.1\%. From these we compute, using the scaling relations
$\gamma_s/\nu=2-\eta_s$ and $2\beta_s/\nu=D-2+\eta_s$, the values of
$\eta_s$ with acceptable accuracy (see columns 4 and 5 of table
\ref{EXPONENTES}). In this case, the corrections to scaling are only
significative for the $L=6$ lattice.  We quote as our preferred value
\begin{equation}
\eta_s=0.038(2).
\end{equation}

For the non staggered sector, we observe that the usual susceptibility
diverges much more slowly than the staggered one ($\gamma_s-\gamma\sim
1.02$). The results for $\eta$, using the corresponding scaling
relations, are reported in the last two columns of table
\ref{EXPONENTES}. In this case the corrections to scaling are non
negligible for all lattice sizes. As the data fit very well to a
linear function of $1/L$ we take as the $L\to\infty$ value
\begin{equation}
\eta=1.338(8),
\end{equation}
where the error is half statistical and half due to the possible
deviations from linearity.  Notice that the large value of $\eta$
means that, at the critical point, the spatial correlation function
($G(\mbox{\boldmath $r$})|_{\beta=\beta_c}\sim|\mbox{\boldmath
$r$}|^{-1-\eta}$), in the non staggered sector, decreases much faster
than in the staggered case.

Our results show that the $O(3)$ symmetry is fully broken in the
ordered phase near the critical point. If we discard order $t^\beta$
terms ($t$ being the reduced temperature), ${\bf M}$ is zero and the
eigenvalues of ${\bf M}_s$, are $\{\epsilon,0,-\epsilon\}$ with
$\epsilon\propto t^{\beta_s}$ ($\beta-\beta_s\sim 0.51)$.
Consequently, the magnetization tensors of the even and odd
sublattices are opposite and the eigenvectors corresponding to the
maximum eigenvalues are orthogonal, and similarly for the minimum
ones. Considering the order $t^\beta$ terms the orthogonality will
only hold approximately.

\medskip

We have studied a spin model in three dimensions with the symmetries
of the $O(3)$ group but with very interesting new properties: it
presents an ordered vacuum where the $O(3)$ symmetry is fully broken;
the transition belongs to a new Universality Class, as the thermal
exponent $\nu$ is different from previously known and, finally, the
model has two odd (magnetic type) relevant operators with different
associated $\eta$ exponents. This may explain why previous
perturbative calculations fail to work for this model.

We think that in addition to the interest of the model by itself, the
results suggest further studies of related models, like the addition
of vector interactions, or four dimensional systems.

We thank Alan Sokal for many enlightening discussions at the beginning
of this work, specially regarding the structure of the vacuum in the
$T\to 0$ limit and the Finite Size Scaling techniques. We are also
indebted to Jos\'e Luis Alonso, Juan Jes\'us Ruiz-Lorenzo and Alfonso
Taranc\'on. This work has been partially supported by CICyT
AEN93-0604, AEN95-1284-E and AEN93-0776.  H.~G.~Ballesteros and
V.~Mart\'{\i}n-Mayor are MEC fellows.

\end{document}